\pdfoutput=1

\documentclass[11pt]{article}
\usepackage{acl}

\usepackage{times}
\usepackage{latexsym}
\usepackage{tabularx}
\usepackage{makecell}
\usepackage{graphicx}
\usepackage{multirow}
\usepackage{color,soul}
\usepackage{float}
\usepackage{xspace}
\usepackage{booktabs}
\usepackage{array}
\usepackage[T1]{fontenc}
\usepackage[utf8]{inputenc}
\usepackage{microtype}
\usepackage{inconsolata}
\usepackage[linesnumbered,ruled,vlined]{algorithm2e}
\usepackage{enumitem}
\usepackage{longtable}
\usepackage{amsmath}
\usepackage{cleveref}
\usepackage{pgfplots}
\pgfplotsset{compat=1.18}
\usepackage{subcaption}

\usepackage{xcolor} 
\usepackage{tikz}
\usetikzlibrary{fit,calc}

\colorlet{mypink}{red!40}
\colorlet{myblue}{cyan!60}

\newcommand{\Dataset}{\textbf{SUPERB}\xspace}

\newcommand\blfootnote[1]{
    \begingroup
    \renewcommand\thefootnote{}\footnote{#1}
    \addtocounter{footnote}{-1}
    \endgroup
}

\title{Generative Product Recommendations for Implicit Superlative Queries}

\author{Kaustubh D. Dhole{$^{\alpha*}$}, Nikhita Vedula$^\beta$, Saar Kuzi$^\beta$, Giuseppe Castellucci$^\beta$ \\ 
{\bf Eugene Agichtein$^{\alpha*}$,} {\bf Shervin Malmasi$^\beta$}\\
\textsuperscript{$\alpha$}{Emory University, Atlanta, GA} \hspace{30pt}
\textsuperscript{$\beta$}{Amazon.com Inc., ~ Seattle, WA, USA} \\
\textcolor{darkblue}{\texttt{kdhole@emory.edu}, \texttt{\{veduln,skuzi,giusecas,eugeneag,malmasi\}@amazon.com}}
}

\begin{document}
\maketitle
\begin{abstract}
In Recommender Systems, users often seek the~\textit{best} products through indirect, vague, or under-specified queries, such as ``\textit{best shoes for trail running}''. Such queries, also referred to as \textit{implicit superlative queries}, pose a significant challenge for standard retrieval and ranking systems as they lack an explicit mention of attributes and require identifying and reasoning over complex factors. We investigate how Large Language Models (LLMs) can generate implicit attributes for ranking as well as reason over them to improve product recommendations for such queries. As a first step, we propose a novel four-point schema for annotating the best product candidates for superlative queries called \Dataset, paired with LLM-based product annotations. We then empirically evaluate several existing retrieval and ranking approaches on our new dataset, providing insights and discussing their integration into real-world e-commerce production systems.
\blfootnote{$^{*}$Work done at Amazon.}
\end{abstract}

\section{Introduction}
\label{sec:intro}

\textit{Superlative queries} are common in product search as users seek products with the highest degree of one or more attributes to satisfy their needs. While some superlative queries can be handled by existing retrieval systems~\cite{kumar2024ranking,zhang2015semantic} through attribute-based filtering (e.g., ``the largest M2 Pro with 32 GB RAM''), others can pose challenges to the existing solutions.
Specifically, in this paper, we study the problem of product ranking and recommendation for~\textit{implicit superlative queries}, where the desired product attributes are not explicitly stated. These queries often involve aspects that require common sense knowledge of the product~\cite{bos2006empirical,scheible-2007-towards}. This problem is further compounded by users creating vague and under specified search queries, either due to a lack of knowledge about certain entity features or the search spanning implicit dimensions, frequently leading to query-product mismatches. For example, a query such as ``\textbf{\textit{the best toy for a 3 year old girl}}'' requires gauging the best products across several implicit attributes. Therefore, to effectively serve such a query, product recommendations should consider popular toy standards like ASTM F963, quality, non-toxic materials, and bright, engaging colors — attributes that are often unknown to end users. With a plethora of product options available on e-commerce platforms, identifying the best products to meet customer needs requires additional product category and world knowledge. Users would then compare multiple products across these attributes before purchasing the best product that fits their needs~\cite{vedula2022matters,vedula2023generating}.

\begin{table}[!t]
\centering
\resizebox{\columnwidth}{!}{%
\begin{tabular}{p{0.3\textwidth}|p{0.23\textwidth}|p{0.35\textwidth}}
\toprule
\textbf{Queries} & \textbf{Query Type} & \textbf{Ranking Criteria} \\ \midrule
\textit{toys} & \small{Expecting Relevant Products} & \small{No Superlative criteria.} \\ \midrule
\textit{highest rated toy for 3-year olds} & \small{Objective Superlative} & \small{\textit{Single Objective Criteria:} highest rating} \\ \midrule
\textit{best toy for my 3-year nephew who loves the Flintstones} & \small{Implicit Superlative} & \small{\textit{Multiple \& Implicit Criteria:} highly-rated, overall positively-reviewed, suitable for a male child, likes Flintstones, dinosaurs, etc.} \\ \bottomrule
\end{tabular}}
\caption{Types of queries along with the criteria of each.~\Dataset{} focuses on implicit superlative queries.}
\label{table:query_types}
\end{table}

Existing ranking pipelines~\cite{reddy2022shopping} rely on traditional relevance labels like `Highly Relevant' vs `Irrelevant' or ESCI (Exact, Substitute, Complement, Irrelevant), and are typically designed for highly objective queries. They do not capture the nuances of product quality and the subjective expectations of ``best'' products for a given need. Even when presented with highly relevant recommendations, users generally rely on a variety of sources, including reviews, blogs, discussion portals, and other online content, to obtain the best. 
In such a scenario, Large Language Models (LLMs) trained on vast amounts of data from diverse sources can act as sources of common-sense and world knowledge. They have been exposed to extensive text sources and have demonstrated success in modeling global opinions in various domains~\cite{santurkar2023whose} and predicting user preferences~\cite{kang2023llms}. LLMs can leverage their knowledge to offer expert insights beyond the basic product descriptions, thereby enabling search and ranking based on external knowledge.

We hypothesize that LLMs possess the capability to perform multi-objective optimization over implicit attributes that match user preferences. Hence, LLMs could play a pivotal role in recommending products for superlative queries by (i) offering comprehensive knowledge across multiple product dimensions, and (ii) addressing the inherent subjectivity associated with such queries.

Our work aims to investigate the research question:~\textbf{\textit{Can LLMs effectively rank and recommend the ``best'' products?}}
To that end, we propose a four-level labeling scheme for superlative queries -- \Dataset with LLM-based annotations, and evaluate retrieval effectiveness across multiple traditional and LLM-based ranking pipelines. 
To our knowledge, this is the first work to explore implicit superlative queries for product recommendation.
Specifically, we make the following contributions:
\begin{itemize}
    \item We investigate the challenges in answering superlative queries, and define a four-level labeling scheme for relevance ratings.
    \item We introduce \Dataset,\footnote{~\url{https://github.com/emory-irlab/SUPERB}} \textbf{Super}latives with~\textbf{B}est relevance annotations, a schema of superlative queries, and pair them with LLM-based annotations using four different ranking approaches i.e.,~\textbf{pointwise},~\textbf{pairwise},~\textbf{listwise} and~\textbf{deliberated} prompting.
    \item We evaluate the retrieval effectiveness of multiple ranking pipelines against~\Dataset{}.
\end{itemize}

Our contributions highlight the importance of addressing superlative queries in recommendation systems, an area that has been largely overlooked. By introducing \Dataset{}, we enable the development and evaluation of systems capable of understanding and fulfilling these high-expectation queries, ultimately improving the user experience and enhancing product recommendations.

\section{Related Work}
\label{sec:related}
We now discuss related work to place our contributions in context.

\subsection{LLMs for Ranking and Recommendation}

LLMs have been successfully applied for ranking and recommendation~\cite{yue2023llamarec}. 
Early pointwise ranking approaches~\cite{monobert} fine-tuned BERT~\cite{devlin2019bert} and T5~\cite{raffel2020exploring} with query-document pairs, and showed improved performance across a variety of benchmarks~\cite{craswell2021ms,beir}. Pointwise approaches~\cite{rankllama} ranked items based on scores predicted for individual documents, while pairwise approaches~\cite{qin-etal-2024-large} prompted models with the query and two documents to compare and rank. Others~\cite{pradeep2023rankvicuna,pradeep2023rankzephyr,sun2023chatgpt} explored a listwise ranking strategy by prompting with a list of documents and generating a ranked list of document IDs. These three paradigms were also examined in the context of recommender systems~\cite{yue2023llamarec}. 

\subsection{LLMs for Relevance Labelling}
After showing promise in predicting searcher preferences~\cite{llmscanpredictsearcher}, LLMs have been extensively used in generating relevance labels~\cite{faggioli2023perspectives,yan2024consolidating,oneshotrelevancelabelling,mehrdad2024large,dhole2024llmjudge,dhole-etal-2025-conqret}. As compared to human evaluation, automated relevance labeling is faster and more scalable.

\subsection{Prompting Approaches}
Apart from standard prompting approaches, deliberative prompting~\cite{li2023deliberate,zheng2024take} approaches like Chain-of-Thought~\cite{wei2022chain} and scaling inference time compute~\cite{snell2024scaling,guo2025deepseek} have successfully improved the performance of LLMs. These methods involve the model generating related information, such as reasoning chains or explanations, to elucidate the reasoning process before arriving at an answer. Our deliberated prompting approach, discussed in Section~\ref{creating_best_rel_annotations} is on similar lines, where we seek to regurgitate implicit attributes so as to make them explicit and help in arriving at the appropriate best label.

\subsection{Superlative Search Queries}
Much of the research related to superlatives has focused on applications in question answering, opinion mining, and sentiment analysis. A recent study~\cite{kumar2024ranking} focused on ranking over objective superlatives where the dimensions to compare against (also referred to as the comparison set~\cite{pyatkin2024superlatives}) are often explicitly provided. However, superlative queries often have implicit, vague and complex dimensions, as we explore in this work.

\section{Implicit Superlative Queries}
We now formalize the type of superlative queries that we seek to address. We define implicit superlative queries as those which (i) \textit{seek the highest degree of one or more attributes or features of a product}; and (ii) \textit{are implicit in nature}. These queries involve preferences which are generally popular, subjective, and not just based on quantifiable attributes. E.g., the superlative query, \textit{``best toy for my 3-year nephew who loves the Flintstones''}, requires an implicit understanding that the user might be looking for a good quality toy which is well-rated and reviewed, reasonably priced, age appropriate, and relates to characters or properties of the show ``The Flintstones''. 
The superlative query \textit{``shoes most suited for marathons''} involves understanding that the user is interested in running shoes and possibly in attributes like cushioning, stability, and durability, which are important for long-distance running. 
Addressing such implicit superlative queries would therefore require (i) inferring key hidden attributes, (ii) world knowledge or a general understanding of concepts, and (iii) being able to reason and compare across different related products and ensure that the necessary attributes are of the highest degree. \Cref{table:query_types} shows a summary and examples of targeted queries.

\section{The \Dataset Relevance Scheme}
We design a novel four-category relevance taxonomy to rank, recommend, and evaluate the retrieved product candidates for superlative queries.
\begin{itemize}[noitemsep]
  \item \textbf{\textit{Overall Best (3)}}: reserved for products that excel across a broad spectrum of parameters including quality, user experience, value for money, innovation, aesthetics, and environmental impact, among others. Products in this category represent the best of what is available in the market, meeting or exceeding all the expected criteria.
  \item \textbf{\textit{Almost Best (2)}}: includes products that perform exceptionally well for most criteria but may fall short in one or a few aspects. These products are generally considered top-tier but lack one or more elements that would elevate them to the Overall Best status.
  \item \textbf{\textit{Relevant But Not the Best (1)}}: captures products that are suitable for certain contexts or specific needs but do not represent the best available option across the board. 
  \item \textbf{\textit{Not Relevant (0)}}: products that do not align well with the user’s query or fail to meet the basic standards expected in their category, making them generally not recommended. 
\end{itemize}

We design such a fine-grained system for multiple reasons. Fine-grained labels have been found to be more advantageous than simplistic binary choices~\cite{zhuang2024beyond}. In addition, they facilitate nuanced evaluations and provide comprehensive feedback. For example, differentiating between~\textbf{Overall Best} and~\textbf{Almost Best} might be less obvious when purchasing standard office supplies, where basic functionality is adequate. However, this distinction becomes essential when selecting infant car seats, where the highest safety and technology standards are vital.

\section{Dataset Construction}
We now describe how we generate superlative queries and pair them with products labeled with annotations from our schema.
\subsection{Creation of Superlative Queries}
For generating superlative queries, we employ the Amazon Shopping Queries dataset~\cite{reddy2022shopping}, which consists of search queries each annotated with up to 40 potential items with ESCI relevance judgements.\footnote{Exact (3), Substitute (2), Complement (1), Irrelevant (0)}

Inspired by LLM-based reformulation approaches~\cite{yang2023zero,dhole2024generative,dhole2024generative2}, we prompt the~\texttt{Claude-Sonnet}~\cite{anthropic2024claude35} LLM with tailored few-shot instructions, to reformulate these shopping queries into their superlative counterparts. We then select queries paired with at least five products with the~\textbf{Exact} ESCI label. We consider all the products of such queries for subsequent \Dataset{} annotations.\footnote{Products of the highest relevance might not necessarily be the~\textbf{Overall Best} option.} 
We generate a total of 35,651 superlative queries from 1,825 original queries. The complete prompt is shown in Appendix \Cref{superlativeprompt} and some of the generated queries are shown in Table~\ref{tab:generatedsuperlative1}. 
\begin{table}[!ht]
\centering
\resizebox{\columnwidth}{!}{%
\begin{tabular}{>{\raggedright\arraybackslash}p{2.5cm}|>{\raggedright\arraybackslash}p{5.4cm}}
    \hline
    \textbf{Query} & \textbf{Superlative Queries} \\ \hline
    \textit{\small{``running shoes''}} & \noindent \textit{\small{``best running shoes for flat feet''} \newline ~~~ \small{``best running shoes for rocky terrain''}} \\ \hline
    \textit{\small{``diaper backpack''}} & \textit{\small{``best diaper backpack for twins''}, \small{``most comfortable diaper backpack for back pain''}} \\ \hline
\end{tabular}}
\caption{Examples of generated superlative queries.}
\label{tab:generatedsuperlative1}
\end{table}
\subsection{Creating Relevance Annotations}
\label{creating_best_rel_annotations}

We adopt four methods for annotating the retrieved product candidates with an LLM:~\textbf{pointwise},~\textbf{pairwise},~\textbf{listwise} and~\textbf{deliberated} prompting. 
In the~\textbf{pointwise} approach, we prompt the model with a superlative query $q$ and the description of a product $p_1$, to generate a single annotation label $b_1$ that corresponds to a category in our schema, along with an explanation $E$ (Eq.~\ref{equ1}).
\begin{equation}
\label{equ1}
    (q, p_1) \rightarrow \mathit{\textbf{M}} \rightarrow b_1 + E 
\end{equation}
In the~\textbf{pairwise} approach, we want the model $\textbf{M}$ to compare a product $p_1$ to another product $p_2$. Hence, we prompt $\textbf{M}$ with the additional description $p_2$ and force it to generate two labels $b_1$ and $b_2$ for both products as shown (Eq.~\ref{equ2}).
\begin{equation}
\label{equ2}
(q, p_1, p_2) \rightarrow \mathit{\textbf{M}} \rightarrow b_1 \, b_2 + E 
\end{equation}
In the~\textbf{listwise} approach, we expand the context to $N-1$ additional products. We hypothesize that providing a context of other products would help the model make accurate judgements in inferring the necessary attributes. Besides, it is more efficient as compared to the pointwise approach as it can process multiple products simultaneously and generate category labels for each (Eq.~\ref{equ3}). 
\begin{equation}
\label{equ3}
(q, p_1, \ldots , p_N) \rightarrow \mathit{\textbf{M}} \rightarrow b_1 \, b_2 \, \ldots \, b_N + E 
\end{equation}

Unlike traditional relevance ranking, where pairwise and listwise evaluations seek to generate a ranked list of document ids, our pairwise and listwise evaluations provide additional information through other products, by expecting the `best' labels for 2 or multiple products. The pairwise and listwise approaches allow gauging the properties of other related product(s) for generating the category label of a product.
We do not explicitly force the model to select the highest category (i.e., Overall Best) in these scenarios. 
\begin{align}
\label{eq1}
q \rightarrow \mathit{\textbf{M}} \rightarrow a_q \\
\label{eq2}
(q, a_q, p_1) \rightarrow \mathit{\textbf{M}} \rightarrow b_1 + E   
\end{align}
We also employ a two-step~\textbf{deliberated prompting} strategy inspired by previous studies~\cite{wei2022chain,li2023deliberate,zheng2024take}, which asks the model to deliberate and reason before generating the final answer. We first generate a set of attributes $a_q$ characterizing the best features of products, and then use them to prompt the model to generate the final taxonomy label (Eq.~\ref{eq1}-\ref{eq2}). These attributes serve as potential dimensions for the model to compare against in the subsequent pointwise step. 
Figure~\ref{fig:modeloutputs} shows an example of the label generation process with deliberated prompting.

\begin{figure}[!bt]
    \centering
    \includegraphics[width=\columnwidth]{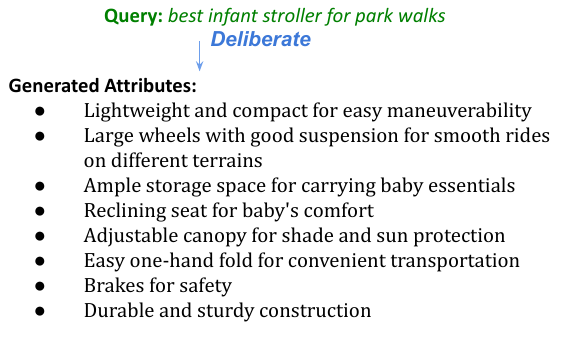}
    \caption{Attributes generated through deliberated prompting for a superlative query.}
    \label{fig:deliberated_generations}
\end{figure}

In e-commerce settings, sellers sometimes exaggerate claims in product descriptions. Making the LLM generate and reason on key useful product attributes prevents the model from providing a biased category judgment based on the seller's description, but instead generate a well-rounded, unbiased categorization in line with the LLM's common sense knowledge.
In each of the methods, we also force the model to generate an explanation to improve model performance~\cite{wei2022chain} and also aid human evaluation. Figure~\ref{fig:deliberated_generations} shows sample generated attributes for a superlative query. We describe the corresponding instructions in Table~\ref{deliberatedprompt} in the Appendix.

\begin{table}[!ht]
\centering
\resizebox{\columnwidth}{!}{%
\begin{tabular}{c|c}
  \hline
  \textbf{Queries} & \textbf{Best Annotations} \\ \hline
  2,230 & 29,218\\ 
  \hline 
  \textbf{Best Label} & \textbf{Number of Examples} \\ \hline
  Overall Best & \phantom{0}8,564 \\
  Almost Best & 10,100 \\
  Relevant But Not the Best & \phantom{0}8,342 \\
  Not Relevant & \phantom{0}2,212 \\
  \hline
\end{tabular}}
\caption{Category label distribution of \Dataset{}.}
\label{tab:label_distribution}
\end{table}

\begin{table*}[!htpb]
\centering
\footnotesize
\resizebox{\textwidth}{!}{%
\begin{tabular}{l|llllll}
\toprule
\textbf{Retrieval Pipeline} & \texttt{P@5} & \texttt{P@10} & \texttt{P@20} & \texttt{nDCG@5} & \texttt{nDCG@10} & \texttt{nDCG@20} \\
\midrule
BM25 & .206 & .163 & .125 & .219 & .213 & .235 \\
RM3 & .214 & .180 & .139 & .219 & .219 & .243 \\ 
\hline
BM25 Top K + Pointwise Reranking & .226 & .163 & -  & .205 & .198 & - \\
RM3\phantom{0} Top K + Pointwise Reranking & .208 & .180 & - 
 & .199 & .210 & - 
 \\
 \hline
BM25 Top K + Listwise Reranking & \textbf{.262}$^\alpha$ & .192$^\alpha$ & .125  & \textbf{.278}$^\alpha$ & \textbf{.259}$^\alpha$ & \textbf{.264}$^\alpha$ \\
RM3\phantom{0} Top K + Listwise Reranking & .248 & \textbf{.201}$^\alpha$ & \textbf{.140} & .245 & .241 & .254 \\

\bottomrule
\end{tabular}%
}
\caption{Performance metrics for different ranking pipelines. $\alpha$ denotes significant improvements via a paired t-test with Holm-Bonferroni~\cite{holm1979simple} correction ($p$-value < 0.05) over BM25. }
\label{tab:performance_metrics}
\end{table*}

\begin{table*}[!ht]
\centering
\begin{tabular}{lcccc}
\hline
\multicolumn{1}{c}{\textbf{Retrieval Pipeline}} & \textbf{P@10} & \textbf{P@50} & \textbf{nDCG@10} & \textbf{nDCG@50} \\
\hline
BM25-Top 100 & .154 & .079 & .205 & .279 \\
BM25-Top 100 + Window (5, 2) & \textbf{.185}$^\alpha$& \textbf{.084}$^\alpha$& \textbf{.241}$^\alpha$& \textbf{.309}$^\alpha$\\
\hline
BM25-Top 100 + Window (20, 10) & .198$^\alpha$& .082$^\alpha$& .240 & .302$^\alpha$\\
BM25-Top 200 & .196 & .079 & .205 & .279 \\
BM25-Top 200 + Window (20, 10) & \textbf{.262} & \textbf{.088} & \textbf{.259} & \textbf{.328} \\
\hline\end{tabular}
\caption{Comparing different retrieval pipelines for the long context setting. $\alpha$ denotes significant improvements via a paired t-test with Holm-Bonferroni~\cite{holm1979simple} correction ($p$-value < 0.05) over BM25.\\}
\label{long_context_results}
\end{table*}

\pagebreak

We use deliberated prompting to generate a large number of \texttt{(query, product, best-label)} triplets, which we refer to as ~\Dataset{}. We generate a total of 29,218 triplets corresponding to 2,230 randomly sampled unique superlative queries. The label distribution is shown in Table~\ref{tab:label_distribution}. Most of the labels are concentrated in the~\textbf{Almost Best} and~\textbf{Relevant But Not the Best} categories, with fewer in the~\textbf{Not Relevant} category. This is expected as annotations were performed over products that were human-rated as~\textbf{Exact}, albeit with respect to the original non-superlative queries. 

\section{Methods}
We perform our analysis in a constrained setting where the item description is limited to 512 tokens in length. This is useful for low latency applications.
We then use \Dataset for evaluating the following ranking pipelines:\\
(i)~\textbf{BM25}: We use BM25 as our baseline.\\
(ii)~\textbf{RM3}: We also employ a pseudo-relevance feedback baseline RM3~\cite{abdul2004umass}.\\
(iii)~\textbf{BM25/RM3 + Listwise Re-ranking}: Here, we re-rank the results of the first stage BM25 and RM3 through a listwise ranking approach. We force the model to generate a ranked list of product IDs in the style of RankGPT~\cite{sun2023chatgpt} (Eq.~\ref{dede}).
    \begin{equation}
    \label{dede}
(q, p_1, \ldots , p_N) \rightarrow \mathit{\textbf{M}} \rightarrow r_1 \, \ldots \, r_N + E 
   \end{equation}
    where \(r_j\) is the index of a product ranked $j$.\\
(iv)~\textbf{BM25/RM3 + Deliberated Pointwise Re-ranking}: Here, the model is forced to generate a schema label for each item along with a confidence score, when given a query and estimated product attributes. The final ranked list is obtained by first sorting using the labels, and resolving ties first by confidence scores, and then by the BM25 scores. This can also be seen as a black-box counterpart of pointwise ranking approaches which provide confidence through logit probabilities. The confidence scores range between 1 and 9 (Eq.~\ref{dede1}-\ref{dede2}).
    \begin{equation}
    \label{dede1}
    q \rightarrow \mathit{\textbf{M}} \rightarrow a_q
    \end{equation}
    \begin{equation}
    \label{dede2}
    (q, p_1, a_q) \rightarrow \mathit{\textbf{M}} \rightarrow b_1 + c_1 + E 
    \end{equation}
(v) \textbf{BM25 + Sliding-Window for long contexts}: We also analyze the case where we use~\textbf{longer product descriptions}, and when there are a~\textbf{large number of products in the context}. In that case, employing a listwise strategy can be detrimental as LLMs have been known to show bias towards specific positions of text in the context~\cite{liu-etal-2024-lost}, while employing a pointwise strategy would involve excessive inference calls. Also, in practice, we observe that LLMs find it hard to generate 100 or 200 item IDs at once, hindering their ability to re-rank items properly. We hence evaluate such queries using a sliding-window approach with BM-25, introduced in RankGPT~\cite{sun2023chatgpt}. 

We choose the~\texttt{Claude-Haiku}~\cite{anthropic2024claude} model for our experiments since it is beneficial to evaluate smaller models for production pipelines. We use the PyTerrier~\cite{pyterrier2020ictir} library with the PyTerrier-GenRank~\cite{dhole2024pyterrier} plugin for designing the retrieval and re-ranking pipelines, and computing~\texttt
{precision (P@k)} and~\texttt{nDCG} metrics. The complete prompts are shown in  Tables~\ref{pointwiseconfidenceprompt} and~\ref{listwiserankingprompt}. We perform a paired t-test with Holm-Bonferroni correction~\cite{holm1979simple} to guage significance.
\\

\section{Results and Analysis}
As shown in Table~\ref{tab:performance_metrics}, we find that the listwise ranking approach is able to rank the best products significantly better as compared to other approaches across all metrics. The listwise scores are better for queries with larger~\texttt{nDCG} values of BM25 meaning they benefit from an initial ranked list as shown in Appendix Figure~\ref{tab:querywise_analysis}. Pointwise approaches also help marginally with~\texttt{P@10} compared to BM25.

We also show the results for top-100 and top-200 items with long descriptions in Table~\ref{long_context_results}. We find that employing a listwise approach in a sliding window fashion significantly improves retrieval effectiveness over the baseline BM25 retrieval across all metrics.
In some cases, we observe modest improvements compared to BM25, highlighting the difficulty of handling superlative queries, which is inherently challenging due to ambiguities and the need for extensive world knowledge. This complexity underscores the hardness of the task for traditional retrieval models. LLMs provide a natural baseline for tackling this issue, given their vast repository of world knowledge, enabling them to approximate knowledge bases and better handle the nuances of superlative queries.

We now examine some error cases in the listwise approach to further explore performance.
By analyzing queries where the methods perform well or poorly, we can gain insights into the model's behavior. The relative performance by nDCG@10 is summarized in Figure~\ref{tab:querywise_analysis} in the Appendix.

\paragraph{Both BM25 and LLM perform well:} Queries like \textit{``most versatile baby carrier for all terrains''} (nDCG@10 of 0.756 and 0.756, respectively) and \textit{``Best of montreal album for summer road trips''} (0.787, 0.951) show strong performance for both approaches. These queries are specific, and the attributes are commonly matched both lexically and semantically to product descriptions.

\paragraph{Both BM25 and LLM perform poorly:} For queries such as \textit{``Most durable kids plates not plastic''} (nDCG@10 of 0.024 and 0.016, respectively) and ``most gentle water wipes for baby’s skin'' (nDCG 0.066 and 0.054, respectively), both approaches struggle. In these cases, challenges like negation, tokenization errors, and specific attributes may contribute to poor performance.

\paragraph{LLM outperforms BM25:}  
Queries like \textit{``most modern LG refrigerators to complement minimalist kitchen decor''} or \textit{``most stylish child safety harness to match toddler’s outfits''} involve interpreting nuances related to style, versatility, and aesthetics, where LLMs arguably excel i.e. recognizing global preferences and broader contexts, enabling them to rerank products with less tangible attributes.

\paragraph{BM25 outperforms LLM:} Many BM25-favored queries have clear, well-defined criteria, such as \textit{``safest bottle warmer for preserving nutrients''} (nDCG 0.508 vs.	0.264); \textit{``most flexible rv caulking sealant for easy application''} (nDCG 0.619 vs. 0.474). We speculate that BM25 excels with queries containing specific product terms and common words as it performs well without advanced reasoning, while LLMs might over-generalize.

\section{Conclusion}
\label{sec:conclusions}
This work studied superlative queries with implicit attributes, which are typically more complex compared to other query types since ranking products for them requires inferring attributes, placing other products in context, and using commonsense and world knowledge to determine the best ones. 
We present \Dataset, a \textbf{4-point schema} to spur further progress in developing product ranking pipelines for superlative queries. We also propose \textbf{pointwise, deliberated pointwise, pairwise, and listwise} methods to label superlative queries over it and re-rank retrieved products, using an LLM as the backbone. 

Our experiments show that LLMs can rank the best items, improve ranking when provided with initial ranked lists, and can also be sensitive to them. In addition, our methods are applicable to rank superlative queries in other item and document ranking settings. The listwise approach is preferable for lower budgets, while the deliberated point-wise approach can be preferred for better quality annotations. We believe that our study can drive further research on superlative search queries.

Our work highlights key considerations for deploying an LLM-based product ranking system into production. While a listwise approach effectively ranks multiple items at once, it can be inefficient due to lengthy item descriptions. In contrast, a pointwise approach is faster, especially with parallel processing. Sliding window methods and query reformulation are also viable alternatives. Generating attributes and explanations clarifies label assignments, boosting user trust and satisfaction.

Addressing superlative queries in product recommendation systems is essential, particularly for the next generation of interactive shopping assistants \cite{vedula:2024,li2025wizardshoppingtargetorientedecommerce} and generative recommender systems \cite{senel2024generative}.
This becomes even more relevant as information-seeking and product search system grow closer together \cite{kuzi:2024}.
These superlative queries capture user intent to find the best possible items, an aspect often overlooked in current systems. Introducing \Dataset allows for the development and assessment of recommendation pipelines capable of handling high-expectation queries, helping systems address this unmet need.

\section*{Limitations}
LLMs have a tendency to average out preferences and often aligning to the majority of the users making them apt for our use case, as shoppers frequently tend to buy the best products unanimously for instance, following viral trends or popular recommendations provided by bloggers. 

However, there are other types of superlative queries that could be subjective and depend on user preferences. It would be interesting to see how such user preferences could be incorporated in ranking the best. We envisage various ways our work could be extended to achieve this -- through traditional techniques like relevance feedback, conversational interactions, and understanding cultural contexts~\cite{dhole2023large,mitchell-etal-2025-shades}. Besides, users often make use of public reviews, blogs and ephemeral trends to guide their purchase decision~\cite{hsu2013effects,wilson2024influence}. Hence incorporating public reviews and other external information through retrieval augmentation could be an interesting line of subsequent study. 

\section*{Acknowledgments}
The authors would like to thank Dhineshkumar Ramasubbu for helping with the annotations, and the anonymous reviewers for their helpful feedback. 

\bibliography{references}

\begin{thebibliography}{52}
\expandafter\ifx\csname natexlab\endcsname\relax\def\natexlab#1{#1}\fi

\bibitem[{Abdul-Jaleel et~al.(2004)Abdul-Jaleel, Allan, Croft, Diaz, Larkey, Li, Smucker, and Wade}]{abdul2004umass}
Nasreen Abdul-Jaleel, James Allan, W~Bruce Croft, Fernando Diaz, Leah Larkey, Xiaoyan Li, Mark~D Smucker, and Courtney Wade. 2004.
\newblock Umass at trec 2004: Novelty and hard.
\newblock \emph{Computer Science Department Faculty Publication Series}, page 189.

\bibitem[{{Anthropic}(2024{\natexlab{a}})}]{anthropic2024claude}
{Anthropic}. 2024{\natexlab{a}}.
\newblock \href {https://www.anthropic.com/news/claude-3-haiku} {Claude 3 haiku: our fastest model yet}.
\newblock Accessed: 2024-07-10.

\bibitem[{{Anthropic}(2024{\natexlab{b}})}]{anthropic2024claude35}
{Anthropic}. 2024{\natexlab{b}}.
\newblock \href {https://www.anthropic.com/news/claude-3-5-sonnet} {Introducing claude 3.5 sonnet}.
\newblock Accessed: 2024-07-10.

\bibitem[{Bos and Nissim(2006)}]{bos2006empirical}
Johan Bos and Malvina Nissim. 2006.
\newblock An empirical approach to the interpretation of superlatives.
\newblock In \emph{Proceedings of the 2006 conference on empirical methods in natural language processing}, pages 9--17.

\bibitem[{Craswell et~al.(2021)Craswell, Mitra, Yilmaz, Campos, and Lin}]{craswell2021ms}
Nick Craswell, Bhaskar Mitra, Emine Yilmaz, Daniel Campos, and Jimmy Lin. 2021.
\newblock Ms marco: Benchmarking ranking models in the large-data regime.
\newblock In \emph{Proceedings of the 44th international ACM SIGIR conference on research and development in information retrieval}, pages 1566--1576.

\bibitem[{Devlin et~al.(2019)Devlin, Chang, Lee, and Toutanova}]{devlin2019bert}
Jacob Devlin, Ming-Wei Chang, Kenton Lee, and Kristina Toutanova. 2019.
\newblock Bert: Pre-training of deep bidirectional transformers for language understanding.
\newblock In \emph{Proceedings of the 2019 Conference of the North American Chapter of the Association for Computational Linguistics: Human Language Technologies, Volume 1 (Long and Short Papers)}, pages 4171--4186.

\bibitem[{Dhole(2023)}]{dhole2023large}
Kaustubh Dhole. 2023.
\newblock Large language models as sociotechnical systems.
\newblock In \emph{Proceedings of the Big Picture Workshop}, pages 66--79.

\bibitem[{Dhole and Agichtein(2024{\natexlab{a}})}]{dhole2024llmjudge}
Kaustubh Dhole and Eugene Agichtein. 2024{\natexlab{a}}.
\newblock \href {https://github.com/emory-irlab/argumentation-rag/blob/main/LLM_Judges_for_Argumentation.pdf} {Llm judges for retrieval augmented argumentation}.

\bibitem[{Dhole(2024)}]{dhole2024pyterrier}
Kaustubh~D Dhole. 2024.
\newblock \href {https://github.com/emory-irlab/pyterrier_genrank} {Pyterrier-genrank: The pyterrier plugin for reranking with large language models}.
\newblock \emph{arXiv preprint arXiv:2412.05339}.

\bibitem[{Dhole and Agichtein(2024{\natexlab{b}})}]{dhole2024generative}
Kaustubh~D Dhole and Eugene Agichtein. 2024{\natexlab{b}}.
\newblock Genqrensemble: Zero-shot llm ensemble prompting for generative query reformulation.
\newblock In \emph{European Conference on Information Retrieval}, pages 326--335. Springer.

\bibitem[{Dhole et~al.(2024)Dhole, Chandradevan, and Agichtein}]{dhole2024generative2}
Kaustubh~D Dhole, Ramraj Chandradevan, and Eugene Agichtein. 2024.
\newblock Generative query reformulation using ensemble prompting, document fusion, and relevance feedback.
\newblock \emph{arXiv preprint arXiv:2405.17658}.

\bibitem[{Dhole et~al.(2025)Dhole, Shu, and Agichtein}]{dhole-etal-2025-conqret}
Kaustubh~D. Dhole, Kai Shu, and Eugene Agichtein. 2025.
\newblock {ConQRet}: Benchmarking fine-grained evaluation of retrieval augmented argumentation with {LLM} judges.
\newblock In \emph{Proceedings of the 2025 Annual Conference of the Nations of the Americas Chapter of the Association for Computational Linguistics}, Albuquerque, New Mexico. Association for Computational Linguistics.

\bibitem[{Faggioli et~al.(2023)Faggioli, Dietz, Clarke, Demartini, Hagen, Hauff, Kando, Kanoulas, Potthast, Stein et~al.}]{faggioli2023perspectives}
Guglielmo Faggioli, Laura Dietz, Charles~LA Clarke, Gianluca Demartini, Matthias Hagen, Claudia Hauff, Noriko Kando, Evangelos Kanoulas, Martin Potthast, Benno Stein, et~al. 2023.
\newblock Perspectives on large language models for relevance judgment.
\newblock In \emph{Proceedings of the 2023 ACM SIGIR International Conference on Theory of Information Retrieval}, pages 39--50.

\bibitem[{Guo et~al.(2025)Guo, Yang, Zhang, Song, Zhang, Xu, Zhu, Ma, Wang, Bi et~al.}]{guo2025deepseek}
Daya Guo, Dejian Yang, Haowei Zhang, Junxiao Song, Ruoyu Zhang, Runxin Xu, Qihao Zhu, Shirong Ma, Peiyi Wang, Xiao Bi, et~al. 2025.
\newblock Deepseek-r1: Incentivizing reasoning capability in llms via reinforcement learning.
\newblock \emph{arXiv preprint arXiv:2501.12948}.

\bibitem[{Holm(1979)}]{holm1979simple}
Sture Holm. 1979.
\newblock A simple sequentially rejective multiple test procedure.
\newblock \emph{Scandinavian journal of statistics}, pages 65--70.

\bibitem[{Hsu et~al.(2013)Hsu, Lin, and Chiang}]{hsu2013effects}
Chin-Lung Hsu, Judy Chuan-Chuan Lin, and Hsiu-Sen Chiang. 2013.
\newblock The effects of blogger recommendations on customers’ online shopping intentions.
\newblock \emph{Internet research}, 23(1):69--88.

\bibitem[{Kang et~al.(2023)Kang, Ni, Mehta, Sathiamoorthy, Hong, Chi, and Cheng}]{kang2023llms}
Wang-Cheng Kang, Jianmo Ni, Nikhil Mehta, Maheswaran Sathiamoorthy, Lichan Hong, Ed~Chi, and Derek~Zhiyuan Cheng. 2023.
\newblock Do llms understand user preferences? evaluating llms on user rating prediction.
\newblock \emph{arXiv preprint arXiv:2305.06474}.

\bibitem[{Kumar et~al.(2024)Kumar, Chatterjee, and Schockaert}]{kumar2024ranking}
Nitesh Kumar, Usashi Chatterjee, and Steven Schockaert. 2024.
\newblock Ranking entities along conceptual space dimensions with llms: An analysis of fine-tuning strategies.
\newblock \emph{arXiv preprint arXiv:2402.15337}.

\bibitem[{Kuzi and Malmasi(2024)}]{kuzi:2024}
Saar Kuzi and Shervin Malmasi. 2024.
\newblock \href {https://doi.org/10.1145/3687273.3687293} {{Bridging the Gap Between Information Seeking and Product Search Systems: Q\&A Recommendation for E-Commerce}}.
\newblock \emph{SIGIR Forum}, 58(1):1–10.

\bibitem[{Li et~al.(2023)Li, Wang, Guo, Song, Tan, Hassan, Menezes, Xiao, Bian, and Zhu}]{li2023deliberate}
Bei Li, Rui Wang, Junliang Guo, Kaitao Song, Xu~Tan, Hany Hassan, Arul Menezes, Tong Xiao, Jiang Bian, and JingBo Zhu. 2023.
\newblock Deliberate then generate: Enhanced prompting framework for text generation.
\newblock \emph{arXiv preprint arXiv:2305.19835}.

\bibitem[{Li et~al.(2025)Li, Chen, Choi, Vedula, Fetahu, Rokhlenko, and Malmasi}]{li2025wizardshoppingtargetorientedecommerce}
Xiangci Li, Zhiyu Chen, Jason~Ingyu Choi, Nikhita Vedula, Besnik Fetahu, Oleg Rokhlenko, and Shervin Malmasi. 2025.
\newblock \href {http://arxiv.org/abs/2502.00969} {Wizard of shopping: Target-oriented e-commerce dialogue generation with decision tree branching}.

\bibitem[{Liu et~al.(2024)Liu, Lin, Hewitt, Paranjape, Bevilacqua, Petroni, and Liang}]{liu-etal-2024-lost}
Nelson~F. Liu, Kevin Lin, John Hewitt, Ashwin Paranjape, Michele Bevilacqua, Fabio Petroni, and Percy Liang. 2024.
\newblock \href {https://doi.org/10.1162/tacl_a_00638} {Lost in the middle: How language models use long contexts}.
\newblock \emph{Transactions of the Association for Computational Linguistics}, 12:157--173.

\bibitem[{Ma et~al.(2024)Ma, Wang, Yang, Wei, and Lin}]{rankllama}
Xueguang Ma, Liang Wang, Nan Yang, Furu Wei, and Jimmy Lin. 2024.
\newblock \href {https://doi.org/10.1145/3626772.3657951} {Fine-tuning llama for multi-stage text retrieval}.
\newblock In \emph{Proceedings of the 47th International ACM SIGIR Conference on Research and Development in Information Retrieval}, SIGIR '24, page 2421–2425, New York, NY, USA. Association for Computing Machinery.

\bibitem[{MacAvaney and Soldaini(2023)}]{oneshotrelevancelabelling}
Sean MacAvaney and Luca Soldaini. 2023.
\newblock \href {https://doi.org/10.1145/3539618.3592032} {One-shot labeling for automatic relevance estimation}.
\newblock In \emph{Proceedings of the 46th International ACM SIGIR Conference on Research and Development in Information Retrieval}, SIGIR '23, page 2230–2235, New York, NY, USA. Association for Computing Machinery.

\bibitem[{Macdonald and Tonellotto(2020)}]{pyterrier2020ictir}
Craig Macdonald and Nicola Tonellotto. 2020.
\newblock Declarative experimentation ininformation retrieval using pyterrier.
\newblock In \emph{Proceedings of ICTIR 2020}.

\bibitem[{Mehrdad et~al.(2024)Mehrdad, Mohapatra, Bagdouri, Chandran, Magnani, Cai, Puthenputhussery, Yadav, Lee, Zhai et~al.}]{mehrdad2024large}
Navid Mehrdad, Hrushikesh Mohapatra, Mossaab Bagdouri, Prijith Chandran, Alessandro Magnani, Xunfan Cai, Ajit Puthenputhussery, Sachin Yadav, Tony Lee, ChengXiang Zhai, et~al. 2024.
\newblock Large language models for relevance judgment in product search.
\newblock \emph{arXiv preprint arXiv:2406.00247}.

\bibitem[{Mitchell et~al.(2025)Mitchell, Attanasio, Baldini, Clinciu, Clive, Delobelle, Dey, Hamilton, Dill, Doughman et~al.}]{mitchell-etal-2025-shades}
Margaret Mitchell, Giuseppe Attanasio, Ioana Baldini, Miruna Clinciu, Jordan Clive, Pieter Delobelle, Manan Dey, Sil Hamilton, Timm Dill, Jad Doughman, et~al. 2025.
\newblock {SHADES}: Towards a multilingual assessment of stereotypes in large language models.
\newblock In \emph{Proceedings of the 2025 Annual Conference of the Nations of the Americas Chapter of the Association for Computational Linguistics}, Mexico City, Mexico. Association for Computational Linguistics.

\bibitem[{Nogueira et~al.(2019)Nogueira, Yang, Cho, and Lin}]{monobert}
Rodrigo Nogueira, Wei Yang, Kyunghyun Cho, and Jimmy Lin. 2019.
\newblock Multi-stage document ranking with bert.
\newblock \emph{arXiv preprint arXiv:1910.14424}.

\bibitem[{Pradeep et~al.(2023{\natexlab{a}})Pradeep, Sharifymoghaddam, and Lin}]{pradeep2023rankvicuna}
Ronak Pradeep, Sahel Sharifymoghaddam, and Jimmy Lin. 2023{\natexlab{a}}.
\newblock Rankvicuna: Zero-shot listwise document reranking with open-source large language models.
\newblock \emph{arXiv preprint arXiv:2309.15088}.

\bibitem[{Pradeep et~al.(2023{\natexlab{b}})Pradeep, Sharifymoghaddam, and Lin}]{pradeep2023rankzephyr}
Ronak Pradeep, Sahel Sharifymoghaddam, and Jimmy Lin. 2023{\natexlab{b}}.
\newblock Rankzephyr: Effective and robust zero-shot listwise reranking is a breeze!
\newblock \emph{arXiv preprint arXiv:2312.02724}.

\bibitem[{Pyatkin et~al.(2024)Pyatkin, Webber, Dagan, and Tsarfaty}]{pyatkin2024superlatives}
Valentina Pyatkin, Bonnie Webber, Ido Dagan, and Reut Tsarfaty. 2024.
\newblock Superlatives in context: Explicit and implicit domain restrictions for superlative frames.
\newblock \emph{arXiv preprint arXiv:2405.20967}.

\bibitem[{Qin et~al.(2024)Qin, Jagerman, Hui, Zhuang, Wu, Yan, Shen, Liu, Liu, Metzler, Wang, and Bendersky}]{qin-etal-2024-large}
Zhen Qin, Rolf Jagerman, Kai Hui, Honglei Zhuang, Junru Wu, Le~Yan, Jiaming Shen, Tianqi Liu, Jialu Liu, Donald Metzler, Xuanhui Wang, and Michael Bendersky. 2024.
\newblock \href {https://aclanthology.org/2024.findings-naacl.97} {Large language models are effective text rankers with pairwise ranking prompting}.
\newblock In \emph{Findings of the Association for Computational Linguistics: NAACL 2024}, pages 1504--1518, Mexico City, Mexico. Association for Computational Linguistics.

\bibitem[{Raffel et~al.(2020)Raffel, Shazeer, Roberts, Lee, Narang, Matena, Zhou, Li, and Liu}]{raffel2020exploring}
Colin Raffel, Noam Shazeer, Adam Roberts, Katherine Lee, Sharan Narang, Michael Matena, Yanqi Zhou, Wei Li, and Peter~J Liu. 2020.
\newblock Exploring the limits of transfer learning with a unified text-to-text transformer.
\newblock \emph{Journal of machine learning research}, 21(140):1--67.

\bibitem[{Reddy et~al.(2022)Reddy, M{\`a}rquez, Valero, Rao, Zaragoza, Bandyopadhyay, Biswas, Xing, and Subbian}]{reddy2022shopping}
Chandan~K Reddy, Llu{\'\i}s M{\`a}rquez, Fran Valero, Nikhil Rao, Hugo Zaragoza, Sambaran Bandyopadhyay, Arnab Biswas, Anlu Xing, and Karthik Subbian. 2022.
\newblock Shopping queries dataset: A large-scale esci benchmark for improving product search.
\newblock \emph{arXiv preprint arXiv:2206.06588}.

\bibitem[{Santurkar et~al.(2023)Santurkar, Durmus, Ladhak, Lee, Liang, and Hashimoto}]{santurkar2023whose}
Shibani Santurkar, Esin Durmus, Faisal Ladhak, Cinoo Lee, Percy Liang, and Tatsunori Hashimoto. 2023.
\newblock Whose opinions do language models reflect?
\newblock In \emph{International Conference on Machine Learning}, pages 29971--30004. PMLR.

\bibitem[{Scheible(2007)}]{scheible-2007-towards}
Silke Scheible. 2007.
\newblock \href {https://aclanthology.org/P07-3012} {Towards a computational treatment of superlatives}.
\newblock In \emph{Proceedings of the {ACL} 2007 Student Research Workshop}, pages 67--72, Prague, Czech Republic. Association for Computational Linguistics.

\bibitem[{Senel et~al.(2024)Senel, Fetahu, Yoshida, Chen, Castellucci, Vedula, Choi, and Malmasi}]{senel2024generative}
L{\"u}tfi~Kerem Senel, Besnik Fetahu, Davis Yoshida, Zhiyu Chen, Giuseppe Castellucci, Nikhita Vedula, Jason Choi, and Shervin Malmasi. 2024.
\newblock \href {https://doi.org/10.18653/v1/2024.acl-long.295} {{Generative Explore-Exploit: Training-free Optimization of Generative Recommender Systems using LLM Optimizers}}.
\newblock In \emph{Proceedings of ACL 2024 (Research Track)}.

\bibitem[{Snell et~al.(2024)Snell, Lee, Xu, and Kumar}]{snell2024scaling}
Charlie Snell, Jaehoon Lee, Kelvin Xu, and Aviral Kumar. 2024.
\newblock Scaling llm test-time compute optimally can be more effective than scaling model parameters.
\newblock \emph{arXiv preprint arXiv:2408.03314}.

\bibitem[{Sun et~al.(2023)Sun, Yan, Ma, Wang, Ren, Chen, Yin, and Ren}]{sun2023chatgpt}
Weiwei Sun, Lingyong Yan, Xinyu Ma, Shuaiqiang Wang, Pengjie Ren, Zhumin Chen, Dawei Yin, and Zhaochun Ren. 2023.
\newblock Is chatgpt good at search? investigating large language models as re-ranking agents.
\newblock In \emph{Proceedings of the 2023 Conference on Empirical Methods in Natural Language Processing}, pages 14918--14937.

\bibitem[{Thakur et~al.(2021)Thakur, Reimers, R{\"u}ckl{\'e}, Srivastava, and Gurevych}]{beir}
Nandan Thakur, Nils Reimers, Andreas R{\"u}ckl{\'e}, Abhishek Srivastava, and Iryna Gurevych. 2021.
\newblock Beir: A heterogeneous benchmark for zero-shot evaluation of information retrieval models.
\newblock \emph{Thirty-fifth Conference on Neural Information Processing Systems Datasets and Benchmarks Track (Round 2)}.

\bibitem[{Thomas et~al.(2024)Thomas, Spielman, Craswell, and Mitra}]{llmscanpredictsearcher}
Paul Thomas, Seth Spielman, Nick Craswell, and Bhaskar Mitra. 2024.
\newblock \href {https://doi.org/10.1145/3626772.3657707} {Large language models can accurately predict searcher preferences}.
\newblock In \emph{Proceedings of the 47th International ACM SIGIR Conference on Research and Development in Information Retrieval}, SIGIR '24, page 1930–1940, New York, NY, USA. Association for Computing Machinery.

\bibitem[{Vedula et~al.(2022)Vedula, Collins, Agichtein, and Rokhlenko}]{vedula2022matters}
Nikhita Vedula, Marcus Collins, Eugene Agichtein, and Oleg Rokhlenko. 2022.
\newblock What matters for shoppers: Investigating key attributes for online product comparison.
\newblock In \emph{European Conference on Information Retrieval}, pages 231--239. Springer.

\bibitem[{Vedula et~al.(2023)Vedula, Collins, Agichtein, and Rokhlenko}]{vedula2023generating}
Nikhita Vedula, Marcus Collins, Eugene Agichtein, and Oleg Rokhlenko. 2023.
\newblock Generating explainable product comparisons for online shopping.
\newblock In \emph{Proceedings of the Sixteenth ACM International Conference on Web Search and Data Mining}, pages 949--957.

\bibitem[{Vedula et~al.(2024)Vedula, Rokhlenko, and Malmasi}]{vedula:2024}
Nikhita Vedula, Oleg Rokhlenko, and Shervin Malmasi. 2024.
\newblock \href {https://doi.org/10.1145/3626772.3661371} {Question suggestion for conversational shopping assistants using product metadata}.
\newblock In \emph{Proceedings of the 47th International ACM SIGIR Conference on Research and Development in Information Retrieval}, SIGIR '24, page 2960–2964, New York, NY, USA. Association for Computing Machinery.

\bibitem[{Wei et~al.(2022)Wei, Wang, Schuurmans, Bosma, Xia, Chi, Le, Zhou et~al.}]{wei2022chain}
Jason Wei, Xuezhi Wang, Dale Schuurmans, Maarten Bosma, Fei Xia, Ed~Chi, Quoc~V Le, Denny Zhou, et~al. 2022.
\newblock Chain-of-thought prompting elicits reasoning in large language models.
\newblock \emph{Advances in neural information processing systems}, 35:24824--24837.

\bibitem[{Wilson et~al.(2024)Wilson, Johnson, and Brown}]{wilson2024influence}
George Wilson, Oliver Johnson, and William Brown. 2024.
\newblock The influence of digital marketing on consumer purchasing decisions.
\newblock \emph{arXiv preprint}.

\bibitem[{Yan et~al.(2024)Yan, Qin, Zhuang, Jagerman, Wang, Bendersky, and Oosterhuis}]{yan2024consolidating}
Le~Yan, Zhen Qin, Honglei Zhuang, Rolf Jagerman, Xuanhui Wang, Michael Bendersky, and Harrie Oosterhuis. 2024.
\newblock Consolidating ranking and relevance predictions of large language models through post-processing.
\newblock \emph{arXiv preprint arXiv:2404.11791}.

\bibitem[{Yang et~al.(2023)Yang, Zhang, and Fang}]{yang2023zero}
Dayu Yang, Yue Zhang, and Hui Fang. 2023.
\newblock Zero-shot query reformulation for conversational search.
\newblock In \emph{Proceedings of the 2023 ACM SIGIR International Conference on Theory of Information Retrieval}, pages 257--263.

\bibitem[{Yue et~al.(2023)Yue, Rabhi, Moreira, Wang, and Oldridge}]{yue2023llamarec}
Zhenrui Yue, Sara Rabhi, Gabriel de Souza~Pereira Moreira, Dong Wang, and Even Oldridge. 2023.
\newblock Llamarec: Two-stage recommendation using large language models for ranking.
\newblock \emph{arXiv preprint arXiv:2311.02089}.

\bibitem[{Zhang et~al.(2015)Zhang, Feng, Huang, Xu, Han, and Zhao}]{zhang2015semantic}
Sheng Zhang, Yansong Feng, Songfang Huang, Kun Xu, Zhe Han, and Dongyan Zhao. 2015.
\newblock Semantic interpretation of superlative expressions via structured knowledge bases.
\newblock In \emph{Proceedings of the 53rd Annual Meeting of the Association for Computational Linguistics and the 7th International Joint Conference on Natural Language Processing (Volume 2: Short Papers)}, pages 225--230.

\bibitem[{Zheng et~al.(2024)Zheng, Mishra, Chen, Cheng, Chi, Le, and Zhou}]{zheng2024take}
Huaixiu~Steven Zheng, Swaroop Mishra, Xinyun Chen, Heng-Tze Cheng, Ed~H. Chi, Quoc~V Le, and Denny Zhou. 2024.
\newblock \href {https://openreview.net/forum?id=3bq3jsvcQ1} {Take a step back: Evoking reasoning via abstraction in large language models}.
\newblock In \emph{The Twelfth International Conference on Learning Representations}.

\bibitem[{Zhuang et~al.(2024)Zhuang, Qin, Hui, Wu, Yan, Wang, and Bendersky}]{zhuang2024beyond}
Honglei Zhuang, Zhen Qin, Kai Hui, Junru Wu, Le~Yan, Xuanhui Wang, and Michael Bendersky. 2024.
\newblock Beyond yes and no: Improving zero-shot llm rankers via scoring fine-grained relevance labels.
\newblock In \emph{Proceedings of the 2024 Conference of the North American Chapter of the Association for Computational Linguistics: Human Language Technologies (Volume 2: Short Papers)}, pages 358--370.

\end{thebibliography}

\newpage

\appendix

\onecolumn

\section*{Appendix}
\label{appendix}

{\small
\begin{center}
\begin{longtable}{|p{15cm}|}
\hline
\label{superlativeprompt}

Given a query, generate multiple diverse superlative versions of the same which require common sense inference. The reformulated superlative queries should provide additional context for which common sense knowledge is required. The context should be related to the item in the original query in various ways and should seek the highest degree of some related aspects. For instance, if a user is looking for a mouse pad, she might be interested in the best one which best complements the color of her laptop, or may require the most suitable one for painful wrists, etc. The context should require generally understood knowledge and common sense and it should not depend on objective criteria like highest rated or cheapest. Some examples of superlative queries are ``Best booster chairs to make mealtime hassle-free for my toddler'', ``most user-friendly diaper pail to make my life as a new mom easier'', ``most suitable lawnmover for rocky areas'', ``most stylish and modern changing table pad to complement my nursery decor'',``Smoothest-riding 2 seater stroller for twin toddlers'',``Best diaper genie for sparking a child's creativity'',``Highest quality epoxy resin for creating stunning wood art pieces'', You should not try to change the type of the product which the user is asking for. Only if the product explicitly mentions a single product, you should change it to make it more generalized (for instance, Amazon \$100 gift card can be changed to \$100 gift card and so on). Do not generate anything else except for one body of JSON and do not explain yourself. Do not include double quotes while generating the superlatives. \\ \\

Provide your output in the form of a JSON.\\ \\

Input Query: LEGO kit \\
\{\{\\
\hspace{1cm}  ``superlatives'' : [\\
\hspace{2cm}      ``best LEGO kit for chess players'',\\
\hspace{2cm}       ``best lego kits for marvel fans'',\\
\hspace{2cm}       ``most impressive lego kits for my friend who is fascinated about India'',\\
\hspace{2cm}       ``best lego kit to encourage my toddler to learn astronomy'',\\
 \hspace{2cm}  ]\\
\}\}\\
\\
Input Query: black halter beaded satin long gowns sequin\\
\{\{\\
\hspace{1cm}   ``superlatives'': [\\
\hspace{2cm}     ``Trendiest black halter beaded satin long gowns with sequins for an Afro-themed fashion parade'',\\
\hspace{2cm}     ``Best halter beaded satin long gowns to match my husband's black silk coat'',\\
\hspace{2cm}     ``Most casual black halter satin long gowns with sequins helpful '',\\
\hspace{2cm}     ``most suitable black halter beaded satin long gowns sequin for a date night''\\
 \hspace{2cm}  ]\\
\}\}\\
\\
Input Query: armani exchange glasses\\
\{\{\\
\hspace{1cm}   ``superlatives'': [\\
\hspace{2cm}     ``best glasses with bold and trendy frames'',\\
\hspace{2cm}     ``best glasses which can be used for office and at parties'',\\
\hspace{2cm}     ``best retro look armani exchange glasses'',\\
\hspace{2cm}     ``most suitable armani exchange glasses for travelling to dubai and mexico" ,\\
\hspace{2cm}     ``best armani exchange glasses that blend seamlessly with my red jeans'',\\
 \hspace{2cm}    ]\\
\}\}\\
\\
Input Query: \{\texttt{query}\}\\
\hline
\caption{Prompt used for Superlative Query Generation}
\end{longtable}
\end{center}
}
\vspace{140pt}
{\small
\begin{center}
\begin{longtable}{|p{15cm}|}
\hline
\label{pointwiseprompt}

Based on the item description and some of its reviews, your internal knowledge about all the features of such types of items, and a user's given shopping query, you should classify the item into one of the taxonomy categories:\\
\\
User Query: \{\texttt{query}\}\\
Item Description: Title: \{\texttt{title}\} Description: \{\texttt{description}\}\\
User Query: \{\texttt{query}\}\\
\\
Categories:\\
3. Overall Best: The item meets the following criteria: The item is overall best in its category on various parameters -- excellence in quality, user experience, value for money, innovation, aesthetics, environmental impact, market position, safety, versatility, processing speed, user rating, etc..\\
2. Almost Best: The item scores high on most or majority of the parameters except for a few. Most users would consider this as item as the best..\\
1. Relevant But Not Best: The item is suitable in certain contexts but not the best option..\\
0. Not Relevant: The item is generally not recommended as it is not relevant to the user's query..\\
\\
Please classify the item into one of the four types. You should return a number between between 3 (Overall Best) and 0 (Not Relevant) followed by an explanation on the next line justifying why that category of best is suitable.\\
\hline
\caption{Pointwise Prompt Used For Best Annotations}
\end{longtable}
\end{center}
}

\begin{figure*}[htpb]
    \centering
    \includegraphics[width=\textwidth]{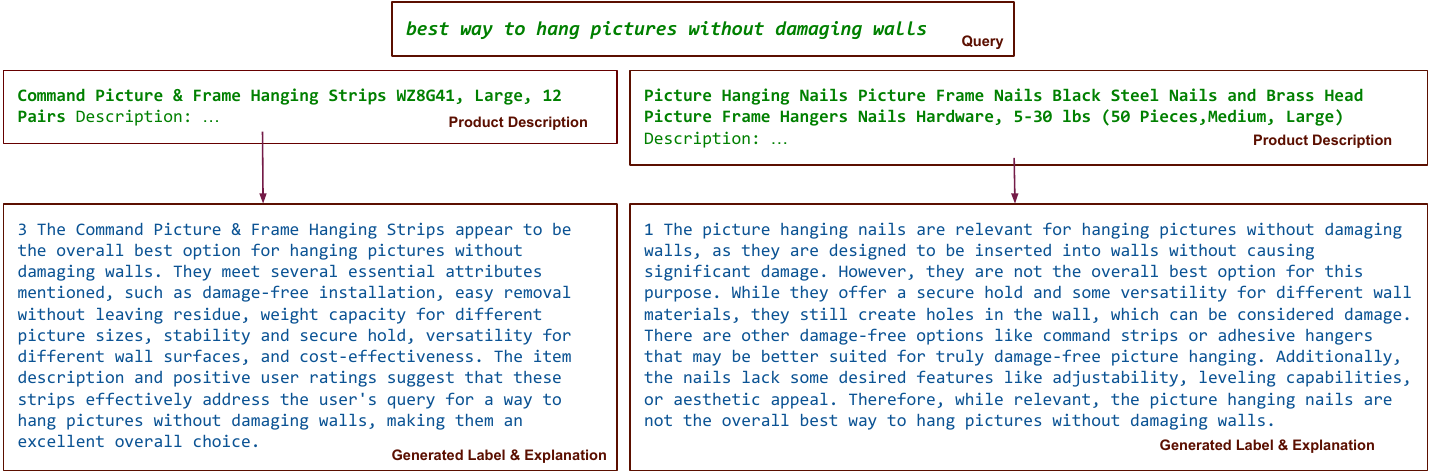} 
    \caption{Sample generated label and explanation using the deliberated pointwise approach.}
    \label{fig:modeloutputs}
\end{figure*}

\section{Evaluating the Best Product Judgements}

To evaluate the efficacy of the~\Dataset{} labels from the above methods, we perform a human evaluation to record the agreement with the model's labels.
In-house domain experts performed the annotation. For each superlative query, the product descriptions, the corresponding category labels and their explanations from the pointwise, pairwise and listwise methods are presented to the annotator, who may agree with none, some, or all of the LLM generated labels. 

As shown in Table~\ref{tab:approval_metrics}, in our first phase of human evaluation, we find that the pointwise approach is more often preferred over listwise and pairwise approaches. During the process of annotation, we find that the pairwise approach tends to narrow its focus on attributes presented in the single product in the context, often misjudging necessary attributes. In the pointwise and listwise approaches, this seems to be less of a concern. 

\begin{table*}[!ht]
\centering
\resizebox{0.5\columnwidth}{!}{%
\begin{tabular}{l|ccc}
\toprule
 & Pointwise & Pairwise & Listwise \\
\midrule
Agreement Rate & \textbf{66.36}\% & 44.86\% & 60.75\% \\
\bottomrule
\end{tabular}}
\caption{Comparing the three best labelling approaches over 107 random superlative queries.}
\label{tab:approval_metrics}
\end{table*}

In the second phase of human evaluation, we use the best strategy of the first phase, i.e., pointwise, and measure the effects of deliberation over a separate set of queries. We find that deliberated prompting is preferred more often than its non-deliberated counterpart, as shown in Table~\ref{tab:approval_metrics_del}, and making the attributes explicit helps assign better quality annotations.

\begin{table}[!ht]
\centering

\resizebox{0.5\columnwidth}{!}{%
\begin{tabular}{l|cc}
\toprule
 & Without & With Deliberation \\
\midrule
Agreement Rate & 75.23\% & \textbf{78.90\%} \\ 
\bottomrule
\end{tabular}}
\caption{Effect of deliberation on pointwise prompting for 109 random superlative queries.}
\label{tab:approval_metrics_del}
\end{table}

\begin{figure*}[ht]
    \centering
    \begin{minipage}{0.5\textwidth}
        \centering
        \includegraphics[width=\columnwidth]{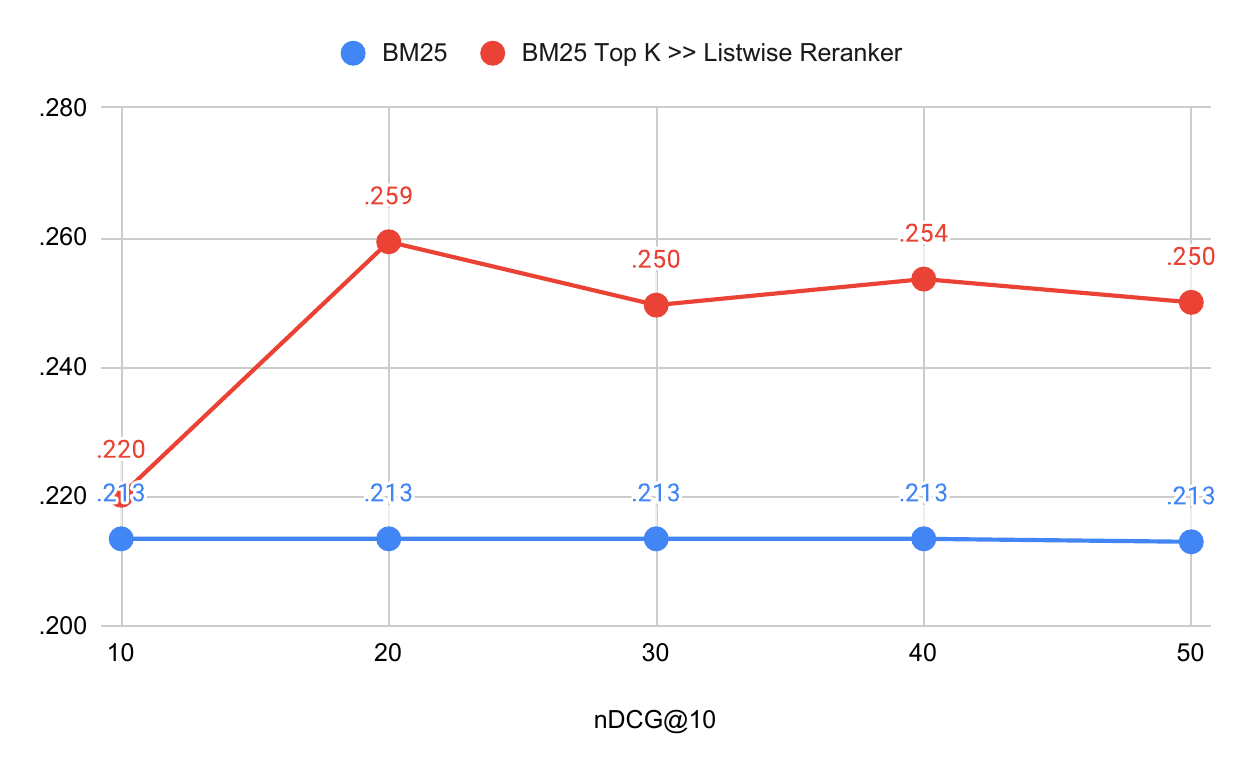}
        \caption{Listwise ranking consistently improves the \\ best ranking for different values of K.}
        \label{fig:figb}
    \end{minipage}%
    \begin{minipage}{0.5\textwidth}
        \centering
        \includegraphics[width=\columnwidth]{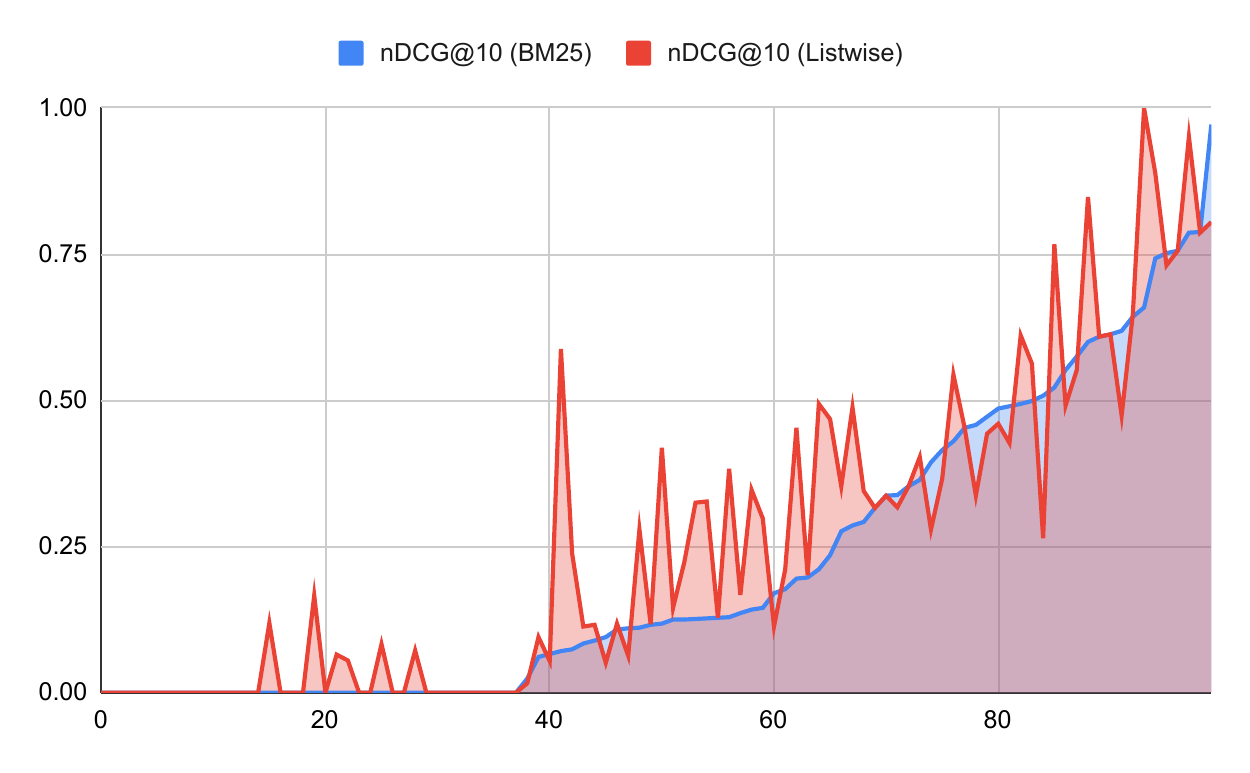}
        \caption{Listwise scores rank better than BM25 for almost all queries. Moreover, LLMs when employed in a listwise fashion benefit from an initial ranked list as queries with higher BM25 scores tend to get better improvements from the listwise approach.}
        \label{tab:querywise_analysis}
    \end{minipage}
\end{figure*}

\textbf{Effect of Increasing the Number of Products:} 
We measure the listwise ranking performance while increasing the number of input products $K$. As shown in Figure~\ref{fig:figb}, we find that the listwise approach increases the likelihood of picking the best product as we provide more products in the context, and then tends to stagnate after a large $K$. The pointwise approach's performance remains almost the same.

As shown in Table~\ref{tab:shufflingorder}, we also shuffle the product order from the first stage retriever and evaluate how sensitive the listwise re-ranker is to the initial order. Shuffling the top-20 products in three different random orders causes drastic performance drops in each, i.e. listwise re-ranking benefits from an initial ranked list and improves upon it.

\begin{table}[!htpb]
\centering
\resizebox{0.5\columnwidth}{!}{%
\begin{tabular}{l|lllll}
\toprule
Listwise & BM25 & \texttt{seed1} & \texttt{seed2} & \texttt{seed3} & RM3\\
\midrule
\texttt{nDCG@10} & .259 & .147 & .143 & .141 & .241\\
\bottomrule 
\end{tabular}
}%
\caption{Listwise reranking performance when the top-20 products are placed in context with initial rankings from BM25, random and RM3 orderings. The listwise re-ranker is highly sensitive to the order provided by the first stage retriever.}
\label{tab:shufflingorder}
\end{table}

\section{Effect of Query Reformulation}
To reduce inference latency for such scenarios, we also investigate incorporating LLM-based reformulation i.e. employing the LLM during query generation rather than during reranking. Specifically, we introduce two types of~\textbf{query reformulations} to generate (i)~\textit{\textbf{keywords}}: this is accomplished by generating generic query expansion terms which are related to the query (ii)~\textit{\textbf{attributes}}: we use the above estimated ideal attributes for expanding the query.

\textbf{Results:} We find that employing keyword and attribute-based reformulated queries helps improve overall retrieval effectiveness, as compared to the original queries. Attribute-based reformulation improves recall and MAP across all retrieval settings.

We find that by employing keyword and attribute based reformulated queries helps improve overall retrieval effectiveness, as compared to the original queries. Attribute based reformulation improves recall and MAP across all retrieval settings. Table~\ref{tab:qr_results} presents the details.
\begin{table*}[ht]
\centering
\begin{tabular}{l|cccccc}
\toprule
\hline
 & \multicolumn{3}{c}{BM25}&  \multicolumn{3}{c}{BM25 + Window (20,10)}\\
\toprule\hline
Queries  &MAP &R@50& \textbf{nDCG@50}  &  MAP &R@50&\textbf{nDCG@50}  \\
\hline
SUPERB (Raw)  &.152 &.358& .279  &  .168&.372&.302\\
+ Keyword based QR  &.155 &.371& \textbf{.291}  &  .172&.383&.31\\
+ Attribute based QR  &~\textbf{.156} &~\textbf{.382}& \textbf{.291}  &~\textbf{.176}&~\textbf{.389}&~\textbf{.311}\\
\hline
\end{tabular}
\caption{Comparison of Query Reformulation with BM25 over superlative queries.}
\label{tab:qr_results}
\end{table*}

{\small
\begin{center}
\begin{longtable}{|p{15cm}|}
\hline
\label{listwiseprompt}

Based on the following descriptions of multiple items and a user's shopping query, you need to classify each item into one of the taxonomy categories:\\
\\
User Query: \{\texttt{query}\}\\
Item 1 Description: Title: \{\texttt{Title 1}\} Description: \{\texttt{Item Description 1}\}\\
Item 2 Description: Title: \{\texttt{Title 2}\} Description: \{\texttt{Item Description 2}\}\\
...\\
...\\
Item N-1 Description: Title: \{\texttt{Title N-1}\} Description: \{\texttt{Item Description N-1}\}\\
Item N Description: Title: \{\texttt{Title N}\} Description: \{\texttt{Item Description N}\}\\
User Query: \{\texttt{query}\}\\
\\
Classification Categories:\\
3. Overall Best: The item meets the following criteria: The item is overall best in its category on various parameters -- excellence in quality, user experience, value for money, innovation, aesthetics, environmental impact, market position, safety, versatility, processing speed, has been rated highly, etc..\\
2. Almost Best: The item scores high on most or majority of the parameters except for a few. Most users would consider this as item as the best..\\
1. Relevant But Not Best: The item is suitable in certain contexts but not the best option..\\
0. Not Relevant: The item is generally not recommended as it is not relevant to the user's query..\\
\\
Please rank each item into one of the four types. First, return the rankings as numbers separated by ' ' where each number ranges between between 3 (Overall Best) and 0 (Not Relevant). And then provide a short explanation as to why you assigned the best categories. You should start your answer with only the rankings (i.e. 3 2 2 0 and so on ) and not a description. Ensure that the number of rankings is equal to the number of items shown i.e. exactly 25.\\
\hline
\caption{Listwise Prompt Used For \Dataset{} Annotations. It provides multiple additional items as context.}
\end{longtable}
\end{center}
}
\vspace{30pt}
{\small
\begin{center}
\begin{longtable}{|p{15cm}|}\hline
\label{deliberatedprompt}
Given a user seeking the best item, define the ideal requirements for satisfying the user query by returning a list of attributes which are essential for that item. For instance, if the user is seeking the best laptop for his 15 year old son, the attributes could be a large RAM, the best GPUs (maybe from NVIDIA or AMD), good speakers etc. You should try to come up attributes which are essential for the perfect or the best item as well as which satisfy the user query. Return your output as a json. Do not generate anything else. 
\{\texttt{query}\}\\
\hline
\caption{Deliberation Step used for Generating Attributes}\\
\end{longtable}
\end{center}

\newpage
\vspace*{-1.5cm}

{\small
\begin{center}
\begin{longtable}{|p{15cm}|}
\hline 
\label{deliberation-eprompt}
Based on the following descriptions of two items, their reviews, and a user's shopping query, you need to rank each item into one of the taxonomy categories:\\
\\
User Query: \{\texttt{query}\}\\
Item 1 Description: Title: \{\texttt{Title 1}\} Description: \{\texttt{Item Description 1}\}\\
Item 2 Description: Title: \{\texttt{Title 2}\} Description: \{\texttt{Item Description 2}\}\\
User Query: \{\texttt{query}\}\\
\\
Categories:\\
3. Overall Best: The item meets the following criteria: The item is overall best in its category on various parameters -- excellence in quality, user experience, value for money, innovation, aesthetics, environmental impact, market position, safety, versatility, processing speed, has been rated highly, etc..\\
2. Almost Best: The item scores high on most or majority of the parameters except for a few. Most users would consider this as item as the best..\\
1. Relevant But Not Best: The item is suitable in certain contexts but not the best option..\\
0. Not Relevant: The item is generally not recommended as it is not relevant to the user's query..\\
\\
Please rank each item into one of the four types. First, return two numbers separated by ' ' where each number ranges between between 3 (Overall Best) and 0 (Not Relevant). And then briefly explain why the category of best is suitable.\\
\hline
\caption{Pairwise Prompt Used For Best Annotations -- Provides one additional item as context}
\end{longtable}
\end{center}
}
\vspace{-15pt}

{\small
\begin{center}
\begin{longtable}{|p{15cm}|}
\hline
\label{pairwiseprompt}
Based on the following descriptions of two items, their reviews, and a user's shopping query, you need to rank each item into one of the taxonomy categories:

User Query: \{\texttt{query}\}\\
The best item would possibly possess many of such attributes: \{Predicted Attributes\}\\
Item 1 Description: Title: \{title\} Description: \{Item Description\}\\
User Query: \{\texttt{query}\}\\
Categories:\\
3. Overall Best: The item meets the following criteria: The item is overall best in its category on various parameters -- excellence in quality, user experience, value for money, innovation, aesthetics, environmental impact, market position, safety, versatility, processing speed, has been rated highly, etc..\\
2. Almost Best: The item scores high on most or majority of the parameters except for a few. Most users would consider this as item as the best..\\
1. Relevant But Not Best: The item is suitable in certain contexts but not the best option..\\
0. Not Relevant: The item is generally not recommended as it is not relevant to the user's query..\\
\\
Please rank each item into one of the four types. First, return two numbers separated by ' ' where each number ranges between between 3 (Overall Best) and 0 (Not Relevant). And then briefly explain why the category of best is suitable.\\
\hline
\caption{Deliberated Pointwise Prompt Used For Best Annotations -- Predicted attributes are provided as context}
\end{longtable}
\end{center}
}
\begin{center}
\begin{longtable}{|p{15cm}|}
\hline
\label{pointwiseconfidenceprompt}
Based on the item description and some of its reviews, your internal knowledge about all the features of such types of items, and a user's given shopping query, you should classify the item into one of the taxonomy categories and provide a confidence score for your prediction:\\
\\
User Query: \{\texttt{query}\}\\
The best item would possibly possess many of such attributes: \{Predicted Attributes\}\\
Item Description: Title: \{\texttt{title}\} Description: \{\texttt{description}\}\\
User Query: \{\texttt{query}\}\\
\\
Categories:\\
3. Overall Best: The item meets the following criteria: The item is overall best in its category on various parameters -- excellence in quality, user experience, value for money, innovation, aesthetics, environmental impact, market position, safety, versatility, processing speed, user rating, etc..\\
2. Almost Best: The item scores high on most or majority of the parameters except for a few. Most users would consider this as item as the best..\\
1. Relevant But Not Best: The item is suitable in certain contexts but not the best option..\\
0. Not Relevant: The item is generally not recommended as it is not relevant to the user's query..\\
\\
You should return a number between between 3 (Overall Best) and 0 (Not Relevant) followed by the confidence of your prediction between 1 to 9 and an explanation on the next line justifying why that category of best is suitable. Your output should look something like this: 2 8 some explanation or 3 4 some explanation. If you are fully confident, then your confidence should have high values like 7, 8 upto 9. If you are not sure, then you should assign low confidence values like 1, 2 or 3. If you are partially confident, then assign other values.\\
\hline
\caption{Deliberated Pointwise Prompt Used For Ranking for generating labels and confidence scores.}
\end{longtable}
\end{center}

\begin{center}
\begin{longtable}{|p{15cm}|}
\hline
\label{listwiserankingprompt}
Based on the following descriptions of multiple items and a user's shopping query, you need to rank the items using the below taxonomy:\\
\\
User Query: \{\texttt{query}\}\\
Item 1 Description: Title: \{\texttt{Title 1}\} Description: \{\texttt{Item Description 1}\}\\
Item 2 Description: Title: \{\texttt{Title 2}\} Description: \{\texttt{Item Description 2}\}\\
...\\
...\\
Item N-1 Description: Title: \{\texttt{Title N-1}\} Description: \{\texttt{Item Description N-1}\}\\
Item N Description: Title: \{\texttt{Title N}\} Description: \{\texttt{Item Description N}\}\\
User Query: \{\texttt{query}\}\\
\\
Classification Categories:\\
3. Overall Best: The item meets the following criteria: The item is overall best in its category on various parameters -- excellence in quality, user experience, value for money, innovation, aesthetics, environmental impact, market position, safety, versatility, processing speed, has been rated highly, etc..\\
2. Almost Best: The item scores high on most or majority of the parameters except for a few. Most users would consider this as item as the best..\\
1. Relevant But Not Best: The item is suitable in certain contexts but not the best option..\\
0. Not Relevant: The item is generally not recommended as it is not relevant to the user's query..\\
\\
The 'Overall Best' item(s) should be ranked higher, followed by the 'Almost Best' item(s), the 'Relevant But not the best' and then the 'not relevant' ones. You should return the item ids separated by ' ' something like 8 3 9 1 2... You should start your answer with only the rankings and not a description. Ensure that each item id is present in the list. Ensure that the number of rankings is equal to the number of items shown i.e. exactly {\texttt{$K$}}.\\
\hline
\caption{Listwise Prompt Used For Ranking}
\end{longtable}
\end{center}
}

\end{document}